\documentclass[aps,prl,twocolumn,amsmath,amssymb,superscriptaddress]{revtex4-1}
\usepackage{graphicx}
\usepackage{dcolumn}
\usepackage{bm}
\usepackage{color}

\begin{document}


\title{Helicoidal Fields and Spin Polarized Currents in CNT-DNA Hybrids}
\author{G. S. Diniz}
\email{ginetom@gmail.com}
\affiliation{Department of Physics and Astronomy and Nanoscale and Quantum Phenomena Institute, Ohio University, Athens, OH 45701, USA}
\author{A. Latg\'e}
\affiliation{Instituto de F\'isica, Universidade Federal Fluminense, Niter\'oi-RJ 24210-340, Brazil}
\author{S. E. Ulloa}
\affiliation{Department of Physics and Astronomy and Nanoscale and Quantum Phenomena Institute, Ohio University, Athens, OH 45701, USA}


\date{\today}

\begin{abstract}
We report on theoretical studies of electronic transport in the archetypical molecular hybrid formed by DNA wrapped around single-walled carbon nanotubes (CNTs). Using a Green's function formalism in a $\pi$-orbital tight-binding representation, we investigate the role that spin-orbit interactions play on the CNT in the case of the helicoidal electric field induced by the polar nature of the adsorbed DNA molecule. We find that spin polarization of the current can take place in the absence of magnetic fields, depending strongly on the direction of the wrapping and length of the helicoidal field.  These findings open new routes for using CNTs in spintronic devices.
\end{abstract}


\maketitle

Carbon nanotubes continue to attract a great deal of attention due to their potential application in a wide variety of electronic devices, including molecule detectors \cite{Meng} and spin-based electronics \cite{Weiss}, to name but a few. Advances in synthesis allow production of CNTs with specific metallic or semiconductor behavior, depending on the tube chirality, and further tailoring of their electronic and optical characteristics via chemical functionalization with specific atoms and molecules \cite{Rubio}.

Many efforts in a variety of carbon nanostructures have been directed to study the effects of spin-orbit interaction  \cite{Ando,Chico2,Huertas,Zhou,Jeong,Jespersen,Loss}. Although this effect is expected to be weak, due to the low atomic number of carbon,
it has been demonstrated in a variety of systems that depending on substrates or adsorbed atoms, the spin-orbit interaction (SOI) can be effectively enhanced in carbon nanostructures \cite{Neto,Rader}. In CNTs in particular, the curvature arising from their cylindrical geometry can enhance SOI via orbital hybridization among neighboring atoms \cite{Chico2,Jeong,Sancho,Loss}.  The application of external electric fields gives rise to Rashba SOI (RSOI) effects that may even create electronic `helical' states for fields applied transversely to the CNT \cite{Loss}. Most significantly, recent experiments have demonstrated that SOI can be quite important in determining the electronic structure and transport properties of CNTs \cite{Kuemmeth}.

Chemical functionalization of CNTs is an important route to manipulate their properties \cite{Burghard,Collins,Fufang}. Prominent among these is the hybrid formed by DNA molecules adsorbed on the exterior walls of the CNT, and found to arrange in a wrapped-around fashion in clock- and counterclockwise helices with pitch (coiling period) which correlates with CNT chirality \cite{Zheng,Meng,Meng2,Dzmitry}.  The DNA molecule is stabilized on the CNT by van der Waals forces, driven by the optimal overlap of $\pi$-orbitals of the DNA bases and of the carbon hexagons on the cylindrical nanotube (i.e., optimal $\pi$-$\pi$ stacking), as well as by electrostatic interactions of the phosphate backbone \cite{Meng,Meng2,Dzmitry}.
Interestingly, as the backbone is charged, the wrapping of the molecule generates sizable helicoidal electric fields on the CNT walls.  Helically modulated potentials have been shown to affect the electronic spectrum of CNTs \cite{Michalski,Puller}, however,
the role of the strong RSOI effects generated by these fields has not been considered.

SOI and helicoidal fields have been identified in recent experiments to produce spin selective transport of photoelectrons on DNA molecules on a gold substrate \cite{Naaman}. Theoretical studies explain the high polarization as arising from the RSOI induced by the helicoidal field built in the chiral molecule, as well as by the inherent low-mobility of carriers in DNA \cite{Cuniberti}, in agreement with direct transport measurements on similar systems \cite{Xie2011}.

In this letter, we present results of electronic transport in CNTs in the presence of RSOI originating from the electric field induced by the DNA molecules. We find that as the typical chiral wrapping of the molecules produces a helicoidal field, this results in {\em spin polarization} of the current that increases quadratically with the number of wrappings for experimental estimates of the RSOI field strength. This polarization in the absence of magnetic fields suggest the possible applications of CNT-DNA hybrids as spin polarization devices, as well as the plausible detection of adsorbed molecules on CNT by means of polarized electrical conductance measurements.


The CNT-DNA system is modeled using a $\pi$-orbital orthogonal tight-binding Hamiltonian that includes the effect of RSOI due to the helical DNA molecule \cite{Martino},
\begin{equation}
H=\sum_{i,\sigma}\epsilon_{i\sigma}c_{i\sigma}^{\dagger}c_{i\sigma} + \sum_{{\langle i,j\rangle \\ {\sigma\sigma^{\prime}}}} 
{\cal{O}}_{\sigma\sigma^{\prime}}
c_{i\sigma}^{\dagger} c_{j\sigma^{\prime}} + h.c.,
\label{H1}
\end{equation}
where $c_{i\sigma}^{\dagger}$ creates an electron at site \emph{i} with spin $\sigma$ ($=\uparrow, \downarrow$) and $\epsilon_{i\sigma}$ is the on-site energy, affected by the strong local fields.  ${\cal{O}}_{\sigma\sigma^{\prime}} = t \delta_{\sigma\sigma^{\prime}}+iV_{R}^{ij}( \vec{u}_{ij}\cdot\vec{s})_{\sigma\sigma^{\prime}}$, includes the
nearest neighbor $\pi$-$\pi$ hopping,
$t\approx -2.9$eV, and the Rashba coupling $V_{R}^{ij}$ that depends on the electric field strength at the nearest neighbor atom pair $\langle i,j \rangle$; $\vec{s}$ are the Pauli matrices, and $\vec{u}_{ij}=\hat{z}\times\vec{\eta}_{ij}$, where $\hat{z}$ is the unit vector perpendicular to the CNT's surface and $\vec{\eta}_{ij}$ is the vector connecting $\langle i,j \rangle$ \cite{Martino,Zarea}. For CNTs with small radius there is an additional SOI term arising from the curvature \cite{Huertas,Loss}; in this work we focus solely on the stronger RSOI produced by the electric field induced by the DNA molecules.

We model $V_R^{ij}$ as a constant $V_R$ for those CNT atoms in close proximity to (underneath) the DNA helix, effectively cutting off the long-range of the helicoidal field distribution.  This short-range approximation mimics the screening of the field by the CNT and background substrate, and it does not qualitatively affect our main results and conclusions.
The electrostatic potential also lifts the on-site degeneracies via $\epsilon_{i\sigma}$ in Eq.\ (\ref{H1}) \cite{Latge,Loss}.

\begin{figure}[!h]
\centering
\includegraphics[scale=0.20]{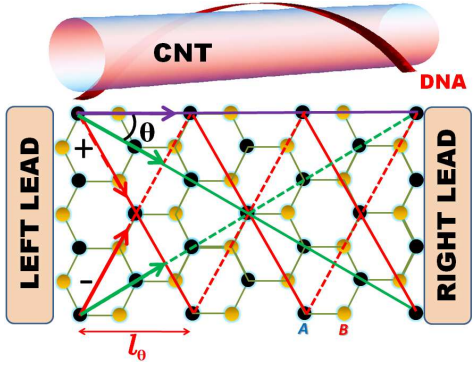}
\caption{(Color online) CNT-DNA hybrid:  DNA wraps around the CNT in a helical fashion. The bottom figure shows an ``unzipped'' version of the CNT system; the central region with adsorbed DNA is connected to left/right semi-infinite CNTs.  DNA helix adsorbs with pitch $l_\theta$, for a given wrapping angle $\theta$ with respect to the tube axis; wrapping can be clockwise (+) or counterclockwise ($-$).  $\theta=0, \pm\pi/6$, and $\pm\pi/3$ are shown.}
\label{scheme}
\end{figure}

The spin-resolved conductance is calculated using a surface Green's function approach in real space \cite{Nardelli}. The CNT device is divided into three regions: left lead, central conductor and right lead, as shown schematically in Fig.\ \ref{scheme}. The central conductor contains the CNT-DNA hybrid, the only region under the influence of RSOI effects; it is also connected to semi-infinite leads by nearest-neighbor hopping with perfect atomic matching, to simulate the DNA adsorbed on a perfect CNT. The Green's function of the central conductor (omitting the spin indices) is then
\begin{equation}
\mathcal G_{C}^{a/r}(E)=\left(\omega_{\pm}-H_{C}-\Sigma_{L}-\Sigma_{R}\right)^{-1},
\end{equation}
where $a/r$ denotes the advanced/retarded Green's function (with energy $\omega_{\pm}=E\pm i\eta$, respectively, $\eta\rightarrow 0$), and $E$ is the energy of the injected electron (the Fermi energy at a given doping). $H_{C}$ stands for the Hamiltonian in the central region and $\Sigma_{L/R}$ are the self-energies for the right/left leads, $\Sigma_{l}=H_{lC}^{\dagger}g_{l}H_{lC}$, where $g_{l}$ is the Green's function for the $l=L,R$ semi-infinite lead, obtained through an iterative procedure \cite{Nardelli}, and $H_{lC}$ couple each lead to the central region. The spin resolved conductance through the entire system is given by,
\begin{equation}
G_{\sigma\sigma^{\prime}}=G_{0}Tr\left[\Gamma_{\sigma}^{L}\mathcal{G}_{C,\sigma\sigma^{\prime}}^{r}\Gamma_{\sigma^{\prime}}^{R}\mathcal{G}_{C,\sigma^{\prime}\sigma}^{a}\right],
\end{equation}
where the trace runs through the lattice sites, $G_{0}=e^2/h$ is the quantum of conductance per spin, and $\Gamma_{\sigma}^{l}$ are the couplings for the leads, related to the spin-diagonal self-energies by $\Gamma^{l}=i\left[\Sigma_{l}^{r}-\Sigma_{l}^{a}\right]$ \cite{Nardelli}.
%

We focus on the results for the electronic conductance of a metallic zigzag (9,0) CNT. These illustrate the general behavior for metallic zigzag tubes, sharing the linear dispersion close to the neutrality point (the Fermi level of undoped CNTs), and similar spin conductances close to the Fermi energy.
For higher (or lower) Fermi energy values (n- or p-doping), the secondary energy shells naturally contribute to the spin conductances for both metallic and semiconductor CNTs, although this requires substantial doping; we focus here on the low-doping behavior.

The wrapping angle $\theta$ of the helical molecule around the CNT is defined so that $\theta=0$ corresponds to a  molecule adsorbed parallel to the CNT long axis. In contrast, a helix with $\theta=\pm\pi/6$ or $\pm\pi/3$ follows the maxima of the orbital wave function on the CNT's surface and produces the expected optimal $\pi$-stacking \cite{Dzmitry}.
As shown below, $\theta$ strongly affects the electronic transport properties on a given nanotube.

\begin{figure}[!h]
\centering
\includegraphics[scale=0.35]{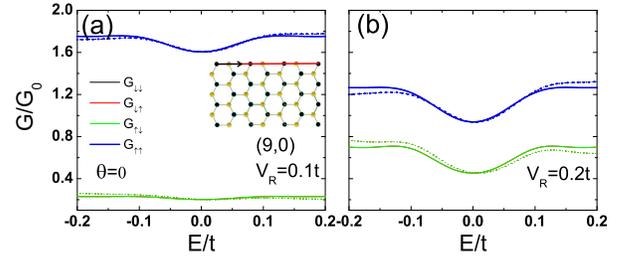}
\caption{(Color online) $G_{\sigma \sigma^\prime}$ for CNT-DNA hybrid with zigzag (9,0) CNT for adsorbed DNA parallel to the CNT axis ($\theta=0$); length of central region is $5.12nm$. (a) Rashba parameter $V_{R}=0.1t$; (b) $V_{R}=0.2t$. Solid lines are for $\epsilon_{i\sigma}=0$; dashed lines $\epsilon_{i\sigma}=0.1t$. Line colors specify $G_{\sigma\sigma^{\prime}}$ in the legend; in all cases $G_{\uparrow\uparrow}=G_{\downarrow\downarrow}$ and $G_{\uparrow\downarrow}=G_{\downarrow\uparrow}$ for this angle $\theta=0$.}
\label{G900degree}
\end{figure}

 Fig.\ \ref{G900degree} shows the conductance spin components for a CNT with adsorbed DNA parallel to the tube axis, i.e. $\theta=0$. The length of the central tube is $4 l_{\theta=\pi/3}=5.12 nm$.  Each panel shows results for different Rashba parameter $V_{R}$, with (solid) and without (dashed) on-site energy modulation. One can identify important features: (i) The diagonal (non-spin-flip) conductance components obey $G_{\uparrow\uparrow}=G_{\downarrow\downarrow}$, while off-diagonal (spin-flip) components are $G_{\downarrow\uparrow}=G_{\uparrow\downarrow}$; these equalities are expected from general considerations for time-reversal and parity preserving fields. (ii) The on-site energy modulation breaks electron-hole symmetry, as the induced electric field acts on the affected sites as donor/acceptor local fields which scatter charge carriers. (iii) As $V_R$ increases, there is a drastic enhancement of the spin-flip channels, as the non-spin-flip conductance components decrease, as one could intuitively expect.  However, notice that symmetry (i) in this case guarantees there is no spin-polarization of the current, regardless of the value of $V_R$ or the length of DNA strand.

\begin{figure}[!h]
\centering
\includegraphics[scale=0.38]{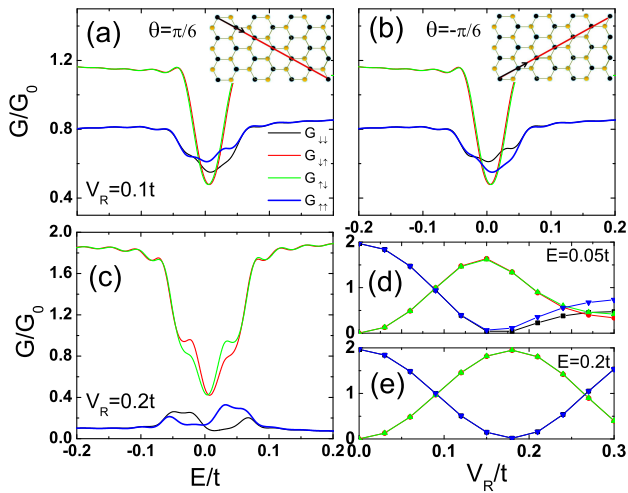}
\caption{(Color online) $G_{\sigma\sigma^{\prime}}$ for (9,0) CNT and DNA  wrapped 4 times with $\theta=\pm \pi/6$, $\epsilon_{i\sigma}=0.1t$.  (a)-(b) Results for $V_{R}=0.1t$. Notice $G_{\uparrow \uparrow}$ exchanges with $G_{\downarrow \downarrow}$ and $G_{\uparrow \downarrow}$ with $G_{\downarrow \uparrow}$  when wrapping direction reverses from (a) to (b). In (c) $V_{R}=0.2t$, $\theta=+\pi/6$.
Colors follow legend in (a). (d)-(e) $G_{\sigma\sigma^{\prime}}$ for $\epsilon_{i\sigma}=0$ and $\theta=+\pi/6$ as function of $V_{R}$ for $E=0.05t$ and $E=0.2t$, respectively.}
\label{G9030degrees}
\end{figure}

Let us now consider the helicoidal field produced by a {\em helical} wrapping of the molecule.  Fig.\ \ref{G9030degrees} shows the conductance spin components when the DNA is wrapped 4 times with $\theta=\pm\pi/6$; each DNA loop extends $l_{\theta=\pi/6}=3.84 nm$ along the tube direction. Panels (a) and (b) show results for the two directions of wrapping, with $V_{R}=0.1t$; (c) shows results for $V_{R}=0.2t$ for the ``+'' direction. (d) and (e) show the oscillatory behavior of the spin components as function of $V_{R}$ for two values of the Fermi energy, $E=0.05t$ and $0.2t$. A few important points are illustrated by these graphs: (i) There is a clear difference between the spin resolved components for both non-spin-flip and  spin-flip conductances, such that $G_{\uparrow \uparrow} \neq G_{\downarrow \downarrow}$, for example. This difference is strongly energy dependent and can be significant for all components.  The splitting of the curves is clearly caused by the chiral parity symmetry breaking of the helicoidal field generated by the wrapped DNA molecule. (ii) The asymmetry exhibited by $G_{\sigma\sigma^{\prime}}$ is reversed when the direction of wrapping reverses. We stress that this asymmetry in  $G_{\sigma \sigma^\prime}$ does not violate time-reversal invariance, as the inversion under wrapping reversal illustrates. In other words, reversing the direction of the current {\em for a given wrapping} results in the reversal of the various $G$ spin components, as expected from symmetry. An important result of this helicoidal field and RSOI is the possibility of spin polarization of the current for a {\em given bias direction}, as we will see below. (iii) The asymmetry in $G_{\sigma \sigma^\prime}$ is not the same for electrons and holes due to the on-site energy modulation considered--this can be easily verified by setting $\epsilon_{i \sigma}=0$. (iv) The different spin components of $G$ depend on an oscillatory manner on the value of $V_R$, as shown in panels (d) and (e), for a given length and wrapping angle of the helicoidal field.  This behavior is reminiscent of the spin field effect transistor and has a similar source \cite{Datta-Das}, as the spin {\em precesses} as it propagates in the presence of the Rashba field, acquiring a net phase that is proportional to $V_RL$, where $L$ is the number of DNA wrappings.
Other wrapping angles show similar behavior.

Results on spin asymmetry and general qualitative dependence on $V_R$ and  wrapping length are similar in other metallic tubes (such as (12,0)), with slightly different energy dependence.  The effects on semiconducting tubes, such as (10,0) only appear, of course, at higher energies (beyond the gap).  Similarly, the values and symmetries among the different spin components of the conductance depend strongly on the wrapping angle.

\begin{figure}[!h]
\centering
\includegraphics[scale=0.35]{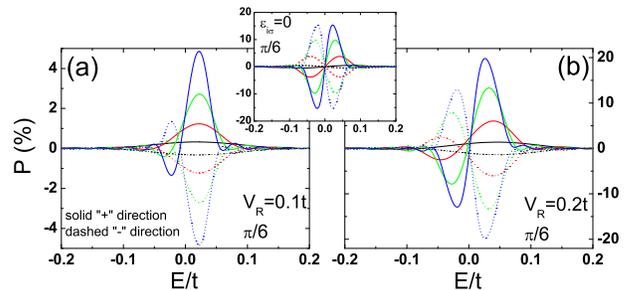}
\caption{(Color online) Normalized conductance polarization $P$ of a (9,0) CNT as function of injected electron energy for different DNA lengths with $\theta =\pm \pi/6$. (a) $V_{R}$ = 0.1t; (b) $V_{R}$ = 0.2t. Solid (dashed) lines
represent results for ``+'' (``$-$'') direction; $P$ increases with number
of wrappings (1 through 4 shown with different color lines).  Central inset shows
$P$ for $\epsilon_{i\sigma}=0$, $V_{R}$ = 0.2t, and $\theta =\pm \pi/6$, displaying full electron-hole asymmetry,
$P(E) = -P(-E)$.}
\label{polar9030degrees}
\end{figure}

We now compute the normalized conductance polarization $P=\sum_{\sigma}(G_{\uparrow\sigma}-G_{\downarrow\sigma})/\sum_{\sigma\sigma^{\prime}}G_{\sigma\sigma^{\prime}}$.
Results for $\theta=\pm\pi/6$ are shown in  Fig.\ \ref{polar9030degrees}  as function of energy.  The polarization is reversed as the wrapping direction is inverted, i.e. as it changes from ``+'' (solid lines) to ``$-$'' (dashed)--a consequence of time-reversal symmetry in the problem.
Notice also that the perfect asymmetry reversal for electrons and holes $P(E)=-P(-E)$ is broken when an on-site energy modulation is present, see inset. The polarization amplitude is strongly dependent on the number of wrappings, growing by nearly one order of magnitude from one full loop (black lines) to four full wrappings (blue). $P$ also increases with $V_{R}$, nearly quadratically in this range, as shown from panel (a) to (b) (notice different vertical scales).

\begin{figure}[!h]
\centering
\includegraphics[scale=0.35]{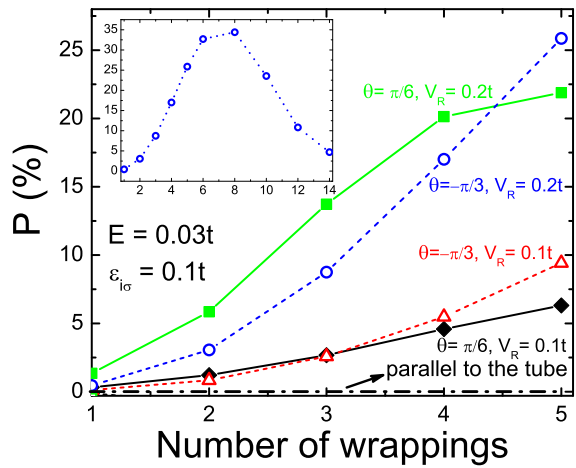}
\caption{(Color online) Normalized polarization $P$ of hybrid with (9,0) CNT as function of number of DNA wrappings and different $V_{R}$ for fixed Fermi energy of E=0.03$t$ and $\epsilon_{i\sigma}=0.1t$. Solid and dashed lines represent results for wrapping angles $\theta=\pi/6$ and $\theta=-\pi/3$, respectively. Dot-dashed black line shows null polarization for the molecule laying parallel to the tube. Reversing direction of wrapping reverses conductance polarization. Inset shows oscillatory behavior of $P$ for $\theta=-\pi/3$, $V_R=0.2t$, for larger range of DNA wrapping.}
\label{helical-length}
\end{figure}

Fig.\ \ref{helical-length} shows $P$ as the number of wrappings $L$ and $V_R$ change, for an energy close to the Fermi energy, $E=0.03t$. DNA parallel to the CNT's axis (dot-dashed line) produces no polarization, regardless of the RSOI strength and length; i.e., a non-helicoidal field produces no spin polarization. On the other hand, non-zero polarization is evident for helicoidal fields. The polarization is found to oscillate for each configuration, as shown in the inset of Fig.\ \ref{helical-length}; the peak polarization value is $\simeq35$\% for $L=8$ for the system with $\theta=-\pi/3$ and $V_{R}=0.2t$.
Similar non-monotonic behavior is found in $P$ for a given $\theta$ and DNA length, as $V_R$ is varied, reaching a peak value similar to that in Fig.\ \ref{helical-length} (not shown).  This oscillatory behavior is in agreement with the expected phase accumulated by the spin precession, and proportional to $V_R L$, as described above. We find that the polarization follows an approximate relation, $P \simeq P_M \sin^2 (V_RL/t )$, with $P_M$ depending on $\theta$ (typically $P_M \simeq 30\%$ to 40\%).  This relation allows one to estimate the polarization for a given CNT-DNA hybrid, once $V_R$ is known.  The strong electric fields induced by the DNA molecule ($\simeq 1.0V/nm$) result in $V_R \simeq 0.2$ to 2 meV (or $10^{-3} t$), in agreement with estimates in the literature \cite{Jespersen,Loss,Kuemmeth}.  For these small $V_R$ values and moderate length, one can then write $P \simeq P_M L^2 /10^6$, which gives a modest polarization of $P \simeq 0.003 \%$ for $L=10$. Additional SOI induced by the CNT curvature is expected to enhance the spin polarization, and may well result in larger $P$ values than these minimal estimates suggest. Measuring this effect would likely require low temperatures in experiments, although increasing the number of wrappings can result in a substantial increase in polarization.

In summary, we find spin polarized conductance for DNA wrapped CNTs due to the presence of helicoidal fields and Rashba spin-orbit coupling--despite the absence of magnetic fields.
This polarization is sensitive to the wrapping pitch and direction, reversing sign for electrons and holes; it can be enhanced by orders of magnitude for a few wrappings. This remarkable behavior can be used in spintronic devices, such as a spin selective filter or in identifying the conformation of molecules adsorbed on these devices.  These results also suggest that other systems with helicoidal Rashba fields would result in conductance polarization without applied magnetic fields, a general result that could in principle be implemented in other electronic systems, such as semiconductor nanowires.

We thank fruitful discussions with M. Zarea, L. Rosales and N. Sandler, and the financial support from Fulbright, CAPES, CNPq and FAPERJ (Brazil) and NSF (DMR and PIRE grants).

\vspace{-2em}
\bibliography{transport-references} 

\begin{thebibliography}{30}%
\makeatletter
\providecommand \@ifxundefined [1]{%
 \@ifx{#1\undefined}
}%
\providecommand \@ifnum [1]{%
 \ifnum #1\expandafter \@firstoftwo
 \else \expandafter \@secondoftwo
 \fi
}%
\providecommand \@ifx [1]{%
 \ifx #1\expandafter \@firstoftwo
 \else \expandafter \@secondoftwo
 \fi
}%
\providecommand \natexlab [1]{#1}%
\providecommand \enquote  [1]{``#1''}%
\providecommand \bibnamefont  [1]{#1}%
\providecommand \bibfnamefont [1]{#1}%
\providecommand \citenamefont [1]{#1}%
\providecommand \href@noop [0]{\@secondoftwo}%
\providecommand \href [0]{\begingroup \@sanitize@url \@href}%
\providecommand \@href[1]{\@@startlink{#1}\@@href}%
\providecommand \@@href[1]{\endgroup#1\@@endlink}%
\providecommand \@sanitize@url [0]{\catcode `\\12\catcode `\$12\catcode
  `\&12\catcode `\#12\catcode `\^12\catcode `\_12\catcode `\%12\relax}%
\providecommand \@@startlink[1]{}%
\providecommand \@@endlink[0]{}%
\providecommand \url  [0]{\begingroup\@sanitize@url \@url }%
\providecommand \@url [1]{\endgroup\@href {#1}{\urlprefix }}%
\providecommand \urlprefix  [0]{URL }%
\providecommand \Eprint [0]{\href }%
\providecommand \doibase [0]{http://dx.doi.org/}%
\providecommand \selectlanguage [0]{\@gobble}%
\providecommand \bibinfo  [0]{\@secondoftwo}%
\providecommand \bibfield  [0]{\@secondoftwo}%
\providecommand \translation [1]{[#1]}%
\providecommand \BibitemOpen [0]{}%
\providecommand \bibitemStop [0]{}%
\providecommand \bibitemNoStop [0]{.\EOS\space}%
\providecommand \EOS [0]{\spacefactor3000\relax}%
\providecommand \BibitemShut  [1]{\csname bibitem#1\endcsname}%
\let\auto@bib@innerbib\@empty
\bibitem [{\citenamefont {Meng}\ \emph
  {et~al.}(2007{\natexlab{a}})\citenamefont {Meng} \emph {et~al.}}]{Meng}%
  \BibitemOpen
  \bibfield  {author} {\bibinfo {author} {\bibfnamefont {S.}~\bibnamefont
  {Meng}} \emph {et~al.},\ }\href {\doibase 10.1021/nl0619103} {\bibfield
  {journal} {\bibinfo  {journal} {Nano Lett.}\ }\textbf {\bibinfo {volume}
  {7}},\ \bibinfo {pages} {45} (\bibinfo {year}
  {2007}{\natexlab{a}})}\BibitemShut {NoStop}%
\bibitem [{\citenamefont {Weiss}\ \emph {et~al.}(2010)\citenamefont {Weiss}
  \emph {et~al.}}]{Weiss}%
  \BibitemOpen
  \bibfield  {author} {\bibinfo {author} {\bibfnamefont {S.}~\bibnamefont
  {Weiss}} \emph {et~al.},\ }\href {\doibase 10.1103/PhysRevB.82.165427}
  {\bibfield  {journal} {\bibinfo  {journal} {Phys. Rev. B}\ }\textbf {\bibinfo
  {volume} {82}},\ \bibinfo {pages} {165427} (\bibinfo {year}
  {2010})}\BibitemShut {NoStop}%
\bibitem [{\citenamefont {Garc\'ia-Lastra}\ \emph {et~al.}(2008)\citenamefont
  {Garc\'ia-Lastra} \emph {et~al.}}]{Rubio}%
  \BibitemOpen
  \bibfield  {author} {\bibinfo {author} {\bibfnamefont {J.~M.}\ \bibnamefont
  {Garc\'ia-Lastra}} \emph {et~al.},\ }\href {\doibase
  10.1103/PhysRevLett.101.236806} {\bibfield  {journal} {\bibinfo  {journal}
  {Phys. Rev. Lett.}\ }\textbf {\bibinfo {volume} {101}},\ \bibinfo {pages}
  {236806} (\bibinfo {year} {2008})}\BibitemShut {NoStop}%
\bibitem [{\citenamefont {Ando}(2000)}]{Ando}%
  \BibitemOpen
  \bibfield  {author} {\bibinfo {author} {\bibfnamefont {T.}~\bibnamefont
  {Ando}},\ }\href {\doibase 10.1143/JPSJ.69.1757} {\bibfield  {journal}
  {\bibinfo  {journal} {J. Phys. Soc. Jpn}\ }\textbf {\bibinfo {volume} {69}},\
  \bibinfo {pages} {1757} (\bibinfo {year} {2000})}\BibitemShut {NoStop}%
\bibitem [{\citenamefont {Chico}\ \emph {et~al.}(2009)\citenamefont {Chico},
  \citenamefont {L\'opez-Sancho},\ and\ \citenamefont {Mu\~noz}}]{Chico2}%
  \BibitemOpen
  \bibfield  {author} {\bibinfo {author} {\bibfnamefont {L.}~\bibnamefont
  {Chico}}, \bibinfo {author} {\bibfnamefont {M.~P.}\ \bibnamefont
  {L\'opez-Sancho}}, \ and\ \bibinfo {author} {\bibfnamefont {M.~C.}\
  \bibnamefont {Mu\~noz}},\ }\href {\doibase 10.1103/PhysRevB.79.235423}
  {\bibfield  {journal} {\bibinfo  {journal} {Phys. Rev. B}\ }\textbf {\bibinfo
  {volume} {79}},\ \bibinfo {pages} {235423} (\bibinfo {year}
  {2009})}\BibitemShut {NoStop}%
\bibitem [{\citenamefont {Huertas-Hernando}\ \emph {et~al.}(2006)\citenamefont
  {Huertas-Hernando}, \citenamefont {Guinea},\ and\ \citenamefont
  {Brataas}}]{Huertas}%
  \BibitemOpen
  \bibfield  {author} {\bibinfo {author} {\bibfnamefont {D.}~\bibnamefont
  {Huertas-Hernando}}, \bibinfo {author} {\bibfnamefont {F.}~\bibnamefont
  {Guinea}}, \ and\ \bibinfo {author} {\bibfnamefont {A.}~\bibnamefont
  {Brataas}},\ }\href {\doibase 10.1103/PhysRevB.74.155426} {\bibfield
  {journal} {\bibinfo  {journal} {Phys. Rev. B}\ }\textbf {\bibinfo {volume}
  {74}},\ \bibinfo {pages} {155426} (\bibinfo {year} {2006})}\BibitemShut
  {NoStop}%
\bibitem [{\citenamefont {Zhou}\ \emph {et~al.}(2009)\citenamefont {Zhou} \emph
  {et~al.}}]{Zhou}%
  \BibitemOpen
  \bibfield  {author} {\bibinfo {author} {\bibfnamefont {J.}~\bibnamefont
  {Zhou}} \emph {et~al.},\ }\href {\doibase 10.1103/PhysRevB.79.195427}
  {\bibfield  {journal} {\bibinfo  {journal} {Phys. Rev. B}\ }\textbf {\bibinfo
  {volume} {79}},\ \bibinfo {pages} {195427} (\bibinfo {year}
  {2009})}\BibitemShut {NoStop}%
\bibitem [{\citenamefont {Jeong}\ and\ \citenamefont {Lee}(2009)}]{Jeong}%
  \BibitemOpen
  \bibfield  {author} {\bibinfo {author} {\bibfnamefont {J.-S.}\ \bibnamefont
  {Jeong}}\ and\ \bibinfo {author} {\bibfnamefont {H.-W.}\ \bibnamefont
  {Lee}},\ }\href {\doibase 10.1103/PhysRevB.80.075409} {\bibfield  {journal}
  {\bibinfo  {journal} {Phys. Rev. B}\ }\textbf {\bibinfo {volume} {80}},\
  \bibinfo {pages} {075409} (\bibinfo {year} {2009})}\BibitemShut {NoStop}%
\bibitem [{\citenamefont {Jespersen}\ \emph {et~al.}(2011)\citenamefont
  {Jespersen} \emph {et~al.}}]{Jespersen}%
  \BibitemOpen
  \bibfield  {author} {\bibinfo {author} {\bibfnamefont {T.~S.}\ \bibnamefont
  {Jespersen}} \emph {et~al.},\ }\href {\doibase 10.1038/nphys1880} {\bibfield
  {journal} {\bibinfo  {journal} {Nat. Phys.}\ }\textbf {\bibinfo {volume}
  {7}},\ \bibinfo {pages} {348} (\bibinfo {year} {2011})}\BibitemShut {NoStop}%
\bibitem [{\citenamefont {Klinovaja}\ \emph {et~al.}(2011)\citenamefont
  {Klinovaja} \emph {et~al.}}]{Loss}%
  \BibitemOpen
  \bibfield  {author} {\bibinfo {author} {\bibfnamefont {J.}~\bibnamefont
  {Klinovaja}} \emph {et~al.},\ }\href {\doibase
  10.1103/PhysRevLett.106.156809} {\bibfield  {journal} {\bibinfo  {journal}
  {Phys. Rev. Lett.}\ }\textbf {\bibinfo {volume} {106}},\ \bibinfo {pages}
  {156809} (\bibinfo {year} {2011})}\BibitemShut {NoStop}%
\bibitem [{\citenamefont {Castro~Neto}\ and\ \citenamefont
  {Guinea}(2009)}]{Neto}%
  \BibitemOpen
  \bibfield  {author} {\bibinfo {author} {\bibfnamefont {A.~H.}\ \bibnamefont
  {Castro~Neto}}\ and\ \bibinfo {author} {\bibfnamefont {F.}~\bibnamefont
  {Guinea}},\ }\href {\doibase 10.1103/PhysRevLett.103.026804} {\bibfield
  {journal} {\bibinfo  {journal} {Phys. Rev. Lett.}\ }\textbf {\bibinfo
  {volume} {103}},\ \bibinfo {pages} {026804} (\bibinfo {year}
  {2009})}\BibitemShut {NoStop}%
\bibitem [{\citenamefont {Varykhalov}\ \emph {et~al.}(2008)\citenamefont
  {Varykhalov} \emph {et~al.}}]{Rader}%
  \BibitemOpen
  \bibfield  {author} {\bibinfo {author} {\bibfnamefont {A.}~\bibnamefont
  {Varykhalov}} \emph {et~al.},\ }\href {\doibase
  10.1103/PhysRevLett.101.157601} {\bibfield  {journal} {\bibinfo  {journal}
  {Phys. Rev. Lett.}\ }\textbf {\bibinfo {volume} {101}},\ \bibinfo {pages}
  {157601} (\bibinfo {year} {2008})}\BibitemShut {NoStop}%
\bibitem [{\citenamefont {L\'opez-Sancho}\ and\ \citenamefont
  {Mu\~noz}(2011)}]{Sancho}%
  \BibitemOpen
  \bibfield  {author} {\bibinfo {author} {\bibfnamefont {M.~P.}\ \bibnamefont
  {L\'opez-Sancho}}\ and\ \bibinfo {author} {\bibfnamefont {M.~C.}\
  \bibnamefont {Mu\~noz}},\ }\href {\doibase 10.1103/PhysRevB.83.075406}
  {\bibfield  {journal} {\bibinfo  {journal} {Phys. Rev. B}\ }\textbf {\bibinfo
  {volume} {83}},\ \bibinfo {pages} {075406} (\bibinfo {year}
  {2011})}\BibitemShut {NoStop}%
\bibitem [{\citenamefont {Kuemmeth}\ \emph {et~al.}(2008)\citenamefont
  {Kuemmeth} \emph {et~al.}}]{Kuemmeth}%
  \BibitemOpen
  \bibfield  {author} {\bibinfo {author} {\bibfnamefont {F.}~\bibnamefont
  {Kuemmeth}} \emph {et~al.},\ }\href {\doibase 10.1038/nature06822} {\bibfield
   {journal} {\bibinfo  {journal} {Nature}\ }\textbf {\bibinfo {volume}
  {452}},\ \bibinfo {pages} {448} (\bibinfo {year} {2008})}\BibitemShut
  {NoStop}%
\bibitem [{\citenamefont {Balasubramanian}\ and\ \citenamefont
  {Burghard}(2005)}]{Burghard}%
  \BibitemOpen
  \bibfield  {author} {\bibinfo {author} {\bibfnamefont {K.}~\bibnamefont
  {Balasubramanian}}\ and\ \bibinfo {author} {\bibfnamefont {M.}~\bibnamefont
  {Burghard}},\ }\href {\doibase 10.1002/smll.200400118} {\bibfield  {journal}
  {\bibinfo  {journal} {Small}\ }\textbf {\bibinfo {volume} {1}},\ \bibinfo
  {pages} {180} (\bibinfo {year} {2005})}\BibitemShut {NoStop}%
\bibitem [{\citenamefont {Collins}\ \emph {et~al.}(2000)\citenamefont {Collins}
  \emph {et~al.}}]{Collins}%
  \BibitemOpen
  \bibfield  {author} {\bibinfo {author} {\bibfnamefont {P.~G.}\ \bibnamefont
  {Collins}} \emph {et~al.},\ }\href {\doibase 10.1126/science.287.5459.1801}
  {\bibfield  {journal} {\bibinfo  {journal} {Science}\ }\textbf {\bibinfo
  {volume} {287}},\ \bibinfo {pages} {1801} (\bibinfo {year}
  {2000})}\BibitemShut {NoStop}%
\bibitem [{\citenamefont {Xu}\ \emph {et~al.}(2007)\citenamefont {Xu} \emph
  {et~al.}}]{Fufang}%
  \BibitemOpen
  \bibfield  {author} {\bibinfo {author} {\bibfnamefont {F.}~\bibnamefont {Xu}}
  \emph {et~al.},\ }\href {\doibase 10.1103/PhysRevB.75.085431} {\bibfield
  {journal} {\bibinfo  {journal} {Phys. Rev. B}\ }\textbf {\bibinfo {volume}
  {75}},\ \bibinfo {pages} {085431} (\bibinfo {year} {2007})}\BibitemShut
  {NoStop}%
\bibitem [{\citenamefont {Zheng}\ \emph {et~al.}(2003)\citenamefont {Zheng}
  \emph {et~al.}}]{Zheng}%
  \BibitemOpen
  \bibfield  {author} {\bibinfo {author} {\bibfnamefont {M.}~\bibnamefont
  {Zheng}} \emph {et~al.},\ }\href {\doibase 10.1126/science.1091911}
  {\bibfield  {journal} {\bibinfo  {journal} {Science}\ }\textbf {\bibinfo
  {volume} {302}},\ \bibinfo {pages} {1545} (\bibinfo {year}
  {2003})}\BibitemShut {NoStop}%
\bibitem [{\citenamefont {Meng}\ \emph
  {et~al.}(2007{\natexlab{b}})\citenamefont {Meng} \emph {et~al.}}]{Meng2}%
  \BibitemOpen
  \bibfield  {author} {\bibinfo {author} {\bibfnamefont {S.}~\bibnamefont
  {Meng}} \emph {et~al.},\ }\href {\doibase 10.1021/nl070953w} {\bibfield
  {journal} {\bibinfo  {journal} {Nano Lett.}\ }\textbf {\bibinfo {volume}
  {7}},\ \bibinfo {pages} {2312} (\bibinfo {year}
  {2007}{\natexlab{b}})}\BibitemShut {NoStop}%
\bibitem [{\citenamefont {Yarotski}\ \emph {et~al.}(2009)\citenamefont
  {Yarotski} \emph {et~al.}}]{Dzmitry}%
  \BibitemOpen
  \bibfield  {author} {\bibinfo {author} {\bibfnamefont {D.~A.}\ \bibnamefont
  {Yarotski}} \emph {et~al.},\ }\href {\doibase 10.1021/nl801455t} {\bibfield
  {journal} {\bibinfo  {journal} {Nano Lett.}\ }\textbf {\bibinfo {volume}
  {9}},\ \bibinfo {pages} {12} (\bibinfo {year} {2009})}\BibitemShut {NoStop}%
\bibitem [{\citenamefont {Michalski}\ and\ \citenamefont
  {Mele}(2008)}]{Michalski}%
  \BibitemOpen
  \bibfield  {author} {\bibinfo {author} {\bibfnamefont {P.~J.}\ \bibnamefont
  {Michalski}}\ and\ \bibinfo {author} {\bibfnamefont {E.~J.}\ \bibnamefont
  {Mele}},\ }\href {\doibase 10.1103/PhysRevB.77.085429} {\bibfield  {journal}
  {\bibinfo  {journal} {Phys. Rev. B}\ }\textbf {\bibinfo {volume} {77}},\
  \bibinfo {pages} {085429} (\bibinfo {year} {2008})}\BibitemShut {NoStop}%
\bibitem [{\citenamefont {{Puller, V. I.}}\ and\ \citenamefont {{Rotkin, S.
  V.}}(2007)}]{Puller}%
  \BibitemOpen
  \bibfield  {author} {\bibinfo {author} {\bibnamefont {{Puller, V. I.}}}\ and\
  \bibinfo {author} {\bibnamefont {{Rotkin, S. V.}}},\ }\href {\doibase
  10.1209/0295-5075/77/27006} {\bibfield  {journal} {\bibinfo  {journal}
  {Europhys. Lett.}\ }\textbf {\bibinfo {volume} {77}},\ \bibinfo {pages}
  {27006} (\bibinfo {year} {2007})}\BibitemShut {NoStop}%
\bibitem [{\citenamefont {G\"ohler}\ \emph {et~al.}(2011)\citenamefont
  {G\"ohler} \emph {et~al.}}]{Naaman}%
  \BibitemOpen
  \bibfield  {author} {\bibinfo {author} {\bibfnamefont {B.}~\bibnamefont
  {G\"ohler}} \emph {et~al.},\ }\href {\doibase 10.1126/science.1199339}
  {\bibfield  {journal} {\bibinfo  {journal} {Science}\ }\textbf {\bibinfo
  {volume} {331}},\ \bibinfo {pages} {894} (\bibinfo {year}
  {2011})}\BibitemShut {NoStop}%
\bibitem [{\citenamefont {Gutierrez}\ \emph {et~al.}(2011)\citenamefont
  {Gutierrez} \emph {et~al.}}]{Cuniberti}%
  \BibitemOpen
  \bibfield  {author} {\bibinfo {author} {\bibfnamefont {R.}~\bibnamefont
  {Gutierrez}} \emph {et~al.},\ }\href {http://arxiv.org/abs/1110.0354}
  {\bibfield  {journal} {\bibinfo  {journal} {arXiv:1110.0354}\ } (\bibinfo
  {year} {2011})}\BibitemShut {NoStop}%
\bibitem [{\citenamefont {Xie}\ \emph {et~al.}(2011)\citenamefont {Xie} \emph
  {et~al.}}]{Xie2011}%
  \BibitemOpen
  \bibfield  {author} {\bibinfo {author} {\bibfnamefont {Z.}~\bibnamefont
  {Xie}} \emph {et~al.},\ }\href@noop {} {\bibfield  {journal} {\bibinfo
  {journal} {Nano Lett.}\ }\textbf {\bibinfo {volume} {11}},\ \bibinfo {pages}
  {4652} (\bibinfo {year} {2011})}\BibitemShut {NoStop}%
\bibitem [{\citenamefont {De~Martino}\ \emph {et~al.}(2002)\citenamefont
  {De~Martino} \emph {et~al.}}]{Martino}%
  \BibitemOpen
  \bibfield  {author} {\bibinfo {author} {\bibfnamefont {A.}~\bibnamefont
  {De~Martino}} \emph {et~al.},\ }\href {\doibase
  10.1103/PhysRevLett.88.206402} {\bibfield  {journal} {\bibinfo  {journal}
  {Phys. Rev. Lett.}\ }\textbf {\bibinfo {volume} {88}},\ \bibinfo {pages}
  {206402} (\bibinfo {year} {2002})}\BibitemShut {NoStop}%
\bibitem [{\citenamefont {Zarea}\ and\ \citenamefont {Sandler}(2009)}]{Zarea}%
  \BibitemOpen
  \bibfield  {author} {\bibinfo {author} {\bibfnamefont {M.}~\bibnamefont
  {Zarea}}\ and\ \bibinfo {author} {\bibfnamefont {N.}~\bibnamefont
  {Sandler}},\ }\href {\doibase 10.1103/PhysRevB.79.165442} {\bibfield
  {journal} {\bibinfo  {journal} {Phys. Rev. B}\ }\textbf {\bibinfo {volume}
  {79}},\ \bibinfo {pages} {165442} (\bibinfo {year} {2009})}\BibitemShut
  {NoStop}%
\bibitem [{\citenamefont {Pacheco}\ \emph {et~al.}(2005)\citenamefont {Pacheco}
  \emph {et~al.}}]{Latge}%
  \BibitemOpen
  \bibfield  {author} {\bibinfo {author} {\bibfnamefont {M.}~\bibnamefont
  {Pacheco}} \emph {et~al.},\ }\href
  {http://stacks.iop.org/0953-8984/17/i=37/a=019} {\bibfield  {journal}
  {\bibinfo  {journal} {J. Phys: Cond. Matt.}\ }\textbf {\bibinfo {volume}
  {17}},\ \bibinfo {pages} {5839} (\bibinfo {year} {2005})}\BibitemShut
  {NoStop}%
\bibitem [{\citenamefont {Nardelli}(1999)}]{Nardelli}%
  \BibitemOpen
  \bibfield  {author} {\bibinfo {author} {\bibfnamefont {M.~B.}\ \bibnamefont
  {Nardelli}},\ }\href {\doibase 10.1103/PhysRevB.60.7828} {\bibfield
  {journal} {\bibinfo  {journal} {Phys. Rev. B}\ }\textbf {\bibinfo {volume}
  {60}},\ \bibinfo {pages} {7828} (\bibinfo {year} {1999})}\BibitemShut
  {NoStop}%
\bibitem [{\citenamefont {Datta}\ and\ \citenamefont {Das}(1990)}]{Datta-Das}%
  \BibitemOpen
  \bibfield  {author} {\bibinfo {author} {\bibfnamefont {S.}~\bibnamefont
  {Datta}}\ and\ \bibinfo {author} {\bibfnamefont {B.}~\bibnamefont {Das}},\
  }\href {\doibase DOI:10.1063/1.102730} {\bibfield  {journal} {\bibinfo
  {journal} {Appl. Phys. Lett.}\ }\textbf {\bibinfo {volume} {56}},\ \bibinfo
  {pages} {665} (\bibinfo {year} {1990})}\BibitemShut {NoStop}%
\end{thebibliography}%

\end{document}